\documentclass[letterpaper,journal]{IEEEtran}

\usepackage[inline]{enumitem}
\usepackage{multicol}
\usepackage{framed} 
\usepackage{cite}
\usepackage{tabularx}
\usepackage{booktabs}

\usepackage{url}
\usepackage{amssymb}
\usepackage{array}
\usepackage{multirow}
\usepackage{amsmath}
\usepackage{graphicx,epstopdf}
\usepackage{svg}
\usepackage{multicol}
\usepackage{float}
\usepackage[ruled,vlined]{algorithm2e}
\usepackage{color}

\usepackage[dvipsnames]{xcolor}  
\definecolor{DarkGreen}{RGB}{0,100,0}  
\usepackage{tikz}
\usetikzlibrary{shapes.geometric, positioning, fit, calc, shapes, arrows, decorations.markings}
\usepackage{subfig}
\usepackage{caption}
\usepackage{threeparttable}

\usepackage{threeparttable}
\usepackage{todonotes}
\usepackage{comment}

\usepackage[inline]{enumitem}
\usepackage{multicol}
\usepackage{framed} 
\usepackage{cite}
\usepackage{tabularx}
\usepackage{booktabs}
\usepackage{multirow} 
\usepackage{cite}
\usepackage{color, soul}

\usepackage{pgf}
\usepackage{tikz}
\usetikzlibrary{positioning, fit, calc, shapes, arrows}
\usepackage{algorithmic}

\PassOptionsToPackage{hyphens}{url}
\usepackage{hyperref}
\hypersetup{hidelinks,hyperfootnotes=false}


\begin{document}
\title{Zero-Knowledge Audit for Internet of Agents: Privacy-Preserving Communication Verification with Model Context Protocol}

\author{Guanlin Jing,
        Huayi Qi
       }

\makeatletter
\let\@IEEEoldmaketitle\maketitle
\renewcommand{\maketitle}{%
  \begingroup
  \def\thefootnote{}%
  \def\footnotemark{}%
  \long\def\@makefnmark##1{}%
  \let\@makefnmark\relax
  \@IEEEoldmaketitle
  \footnotetext{Guanlin Jing, E-mail: guanl001@bjut.edu.cn, Huayi Qi, E-mail: qi@huayi.email}%
  \endgroup
  \def\thefootnote{\@arabic\c@footnote}%
}
\makeatother

\maketitle

\begin{abstract}
        Existing agent communication frameworks face critical limitations in providing verifiable audit trails without compromising the privacy and confidentiality of agent interactions. The protection of agent communication privacy while ensuring auditability emerges as a fundamental challenge for applications requiring accurate billing, compliance verification, and accountability in regulated environments.

        We introduce a framework for auditing agent communications that keeps messages private while still checking they follow expected rules. It pairs zero-knowledge proofs with the existing Model Context Protocol (MCP) so messages can be verified without revealing their contents. The approach runs in lightweight networks, stays compatible with standard MCP exchanges, and adds asynchronous audit verification to confirm format and general message types without exposing specifics.

        The framework enables mutual audits between agents: one side can check communication content and quality while the other verifies usage metrics, all without revealing sensitive information. We formalize security goals and show that zk-MCP provides data authenticity and communication privacy, achieving efficient verification with negligible latency overhead. We fully implement the framework, including Circom-based zero-knowledge proof generation and an audit protocol integrated with MCP's bidirectional channel, and, to our knowledge, this is the first privacy-preserving audit system for agent communications that offers verifiable mutual auditing without exposing message content or compromising agent privacy.
\end{abstract}


\section{Introduction}\label{sec:introduction}

The Internet of Agents (IoA) technology employs autonomous agents, large language models, and advanced communication protocols to realize pervasive network connections and intelligent service exchanges among distributed agents, which gives rise to an integrated system that is capable of dynamic service provisioning, intelligent task coordination, and collaborative problem solving. In addition, IoA has the potential to bring great benefits in improving service efficiency, reducing operational costs, and enabling complex multi-agent collaboration across various domains such as finance, healthcare, and enterprise resource management.

Unfortunately, IoA involves severe security and privacy challenges in addition to the aforementioned great benefits. Current agent communication frameworks, particularly those built on the Model Context Protocol (MCP), excel at facilitating real-time agent coordination and bidirectional message exchange through JSON-RPC 2.0-based communication. However, existing frameworks face critical limitations in their ability to provide verifiable audit trails without compromising the privacy and confidentiality of agent interactions. Most agent service providers collect a large amount of communication data, and employ advanced information technologies such as big data analysis and artificial intelligence to gain valuable knowledge about their clients. This would lead to potential centralization, privacy leakage, and lack of accountability in agent-based systems.

The protection of agent communication privacy while ensuring auditability emerges as a critical problem in recent years. To realize multi-agent collaboration, agents have to periodically exchange service requests, process responses, and consume computational resources (e.g., tokens) during their interactions. Services like financial transaction processing and healthcare data coordination require accurate billing based on actual token consumption and service usage, while also demanding verifiable proof of output authenticity. Without appropriate protection on agent communication privacy, it is easy for adversaries to infer sensitive information through data analytics. This problem becomes even more challenging when audit verification is required for compliance and billing purposes~\cite{narajala2025mcp}. For example, an audit system needs to verify that an agent consumed a certain number of tokens and used a certified model without exposing the specific content of requests or internal computations.

Recently, zero-knowledge proof technology has attracted tremendous interests from government and academia to industry for its ability to prove statement correctness without leaking any additional information~\cite{lavin2024zkp}. As a type of zero-knowledge proof, zero knowledge Succinct Non-interactive ARgument of Knowledge (zk-SNARK) has found its great value in preserving privacy for blockchains such as Zerocash~\cite{bensasson2014zerocash} and BlockMaze~\cite{guan2020blockmaze}. Besides, zk-SNARK's applications are not limited to blockchain systems. It can be employed in more general environments for privacy-preserving audit verification, demonstrating its versatility beyond distributed ledger applications. The MCP, with its streaming characteristics, Server-Sent Events (SSE) support, and HTTP lifecycle constraints, provides an ideal foundation for implementing zero-knowledge verification mechanisms that can ensure integrity while maintaining communication privacy. However, as a privacy-preserving audit system, a delicate mechanism should be designed to ensure verifiable audit trails without compromising the confidentiality of agent communications.

In this paper, we attempt to address this problem and design a zero-knowledge proof-based privacy-preserving audit framework for agent communications in IoA. For the sake of convenience, we highlight our contributions as follows:
\\
1) We design a Zero-Knowledge Agent Communication Audit Framework, a privacy-preserving audit system for agent communications built on the MCP protocol. To the best of our knowledge, this is the first framework that enables verifiable audit verification for agent communications without exposing communication content or compromising agent privacy. A key component within our framework is the integration of zk-SNARK with the MCP protocol, which preserves privacy while guaranteeing correct audit verification through token consumption proof and output authenticity proof. 
\\
2) We formulate a security definition for our zero-knowledge audit framework, and demonstrate its security properties. In addition, we give a comprehensive discussion on practical considerations including the integration with MCP's lifecycle management, capability negotiation, and streaming communication. According to our analysis, one can see that our framework not only can prevent existing attacks targeting agent communication privacy, but also demonstrates high feasibility with verification overhead remaining below 4.14\% of total communication costs.
\\
3) We fully implement our framework including zk-SNARK proof generation using Circom  and the audit protocol that integrates with MCP's bidirectional communication. Our framework leverages MCP's streaming characteristics and HTTP lifecycle constraints to support zero-knowledge verification by providing communication metadata without exposing the actual content. Finally, we conduct comprehensive experiments to evaluate the performance of our framework, and the results show that our framework is highly efficient in processing audit verification while maintaining complete communication privacy.

\textbf{Roadmap.} Sec.~\ref{sec 2} surveys related works on agent communication and zero-knowledge proofs. Sec.~\ref{sec 3} presents the system model and security goals. Sec.~\ref{sec:zk-mcp-protocol} details our zero-knowledge audit framework and protocol. Performance evaluation is shown in Sec.~\ref{sec:performance-experiment}, and the conclusion summarizes our findings in Sec.~\ref{sec:conclusion}.

\section{Related Works}\label{sec 2}
Agent communication protocols have been a fundamental research area in multi-agent systems for decades. Early works on agent communication focused on basic message passing mechanisms and shared blackboard systems~\cite{wooldridge2009introduction}. Subsequent research on agent communication protocols in distributed environments includes~\cite{bellifemine2007developing, fipa2002fipa, odell2001the}, in which communication-efficient protocols using message queues, event-driven architectures, and service-oriented architectures are proposed under reliable communication environments.

\subsection{Model Context Protocol Research}

The MCP has emerged as a transformative paradigm in agent communication, representing a significant advancement in context-aware AI systems. MCP is designed to enable seamless, context-rich communication between Large Language Models (LLMs) and external tools, addressing interoperability and data silo challenges in distributed and multi-agent systems~\cite{hou2025mcp}. Unlike traditional communication protocols that focus solely on message delivery, MCP embeds contextual metadata into data exchanges, enhancing system adaptability, security, and collaborative intelligence across diverse environments.

Recent research on MCP-based agent communication has focused on several key areas. The foundational MCP architecture~\cite{hou2025mcp, patil2025mcp} consists of standardized interfaces and workflows that facilitate the creation, operation, and updating of MCP servers. The protocol embeds metadata such as location, time, and device state into communications, allowing systems to interpret not just data but its context, which is crucial for real-time decision-making and adaptability in dynamic environments. This context-rich approach enables interoperability across LLMs, IoT devices, and cyber-physical systems.

Subsequent research on MCP security and privacy has identified several critical challenges. MCP introduces new security and privacy risks, including potential attack vectors like tool poisoning and unauthorized data access~\cite{hou2025mcp, narajala2025mcp}. Research highlights the need for enterprise-grade mitigation frameworks, including systematic threat modeling, actionable security patterns, and technical controls tailored for MCP implementations. Recommendations emphasize rigorous governance, continuous monitoring, and adaptive security measures throughout the MCP lifecycle.


Additionally, MCP research has explored multi-agent systems and adaptive protocols. The works in~\cite{krishnan2025mcp, vadlamani2025mcp} demonstrate that MCP enables standardized context sharing, scalable coordination, and efficient context management in multi-agent systems. Adaptive protocols within MCP dynamically adjust information exchange based on task complexity and resource constraints, reducing communication overhead while maintaining performance in distributed sensor networks, autonomous vehicles, and collaborative problem-solving scenarios.

However, to the best of our knowledge, no previous work has considered the usage and implementation of zero-knowledge proof-based audit mechanisms in the MCP-based agent communication area.

\subsection{Zero-Knowledge Proofs }

Zero-knowledge proofs (ZKPs) have emerged as a fundamental cryptographic primitive for ensuring privacy and security in distributed systems. ZKPs enable one party to prove the validity of a statement to another without revealing any underlying information, making them particularly valuable for privacy-preserving verification in distributed environments~\cite{motlagh2025zkp, lavin2024zkp}. In the context of agent communication and audit verification, ZKPs offer the potential to verify agent capabilities and tool usage without exposing sensitive communication content.

Recent research on ZKPs in distributed systems has focused on several key areas. The foundational ZKP protocols include zk-SNARKs, zk-STARKs, Bulletproofs, and distributed non-interactive ZKPs, each offering different trade-offs in scalability, efficiency, and proof size for resource-constrained or decentralized environments~\cite{motlagh2025zkp, lavin2024zkp, grilo2025zkp, morais2019zkp}. Distributed ZKP protocols have been developed for classic graph problems and for certifying network properties, with recent advances enabling efficient, communication-minimized proofs across multiple verifiers~\cite{bick2022zkp, grilo2025zkp, boyle2020zkp}.

The applications of ZKPs span multiple domains. In blockchain systems, ZKPs are used for confidential transactions, private smart contracts, and cross-chain privacy protection~\cite{lavin2024zkp, lungu2025zkp, skorobogatova2023zkp}. In cloud and multiparty computation environments, ZKPs verify computations on encrypted data without revealing inputs~\cite{dhokrat2024zkp, zhang2024zkp}. Machine learning and IoT systems also leverage ZKPs for secure, privacy-preserving data sharing and model verification~\cite{lavin2024zkp, kersic2024zkp, zhang2024zkp, yang2023zkp}.

However, the application of ZKPs to agent communication protocols, particularly MCP-based systems, remains largely unexplored. While ZKPs have been successfully applied to blockchain, IoT, and cloud computing scenarios, their integration with context-rich agent communication protocols presents unique challenges. The need to verify agent capabilities, tool usage, and communication integrity while preserving the privacy of contextual metadata requires novel approaches that combine MCP's context-aware communication with ZKP's privacy-preserving verification capabilities.

\subsection{Integration Challenges and Research Gaps}


\begin{table*}[!htb]
\centering
\caption{Related Works on MCP-based Agent Communication and Zero-Knowledge Proofs \label{tab:mcp_comm_protocols}}
\small
\begin{tabular}{|p{2.2cm}|p{3.5cm}|p{7.8cm}|}
\hline \textbf{Reference} & \textbf{Research Area} & \textbf{Detailed descriptions}\\
\hline
\cite{hou2025mcp, patil2025mcp} & \shortstack[l]{MCP Architecture \& \\ Security} & Protocol design, security threats, mitigation strategies\\
\hline
\multirow{2}{*}{\cite{hou2025mcp, narajala2025mcp}} & \multirow{2}{*}{\shortstack[l]{MCP Security Threats \& \\ Mitigation}} & (1) MCP introduces tool poisoning and unauthorized data access risks.\\
& & (2) Enterprise-grade mitigation frameworks with systematic threat modeling.\\
\hline
\multirow{2}{*}{\cite{krishnan2025mcp, vadlamani2025mcp}} & \multirow{2}{*}{\shortstack[l]{Multi-Agent Systems \& \\ Adaptive Protocols}} & (1) MCP enables standardized context sharing and scalable coordination.\\
& & (2) Adaptive protocols dynamically adjust information exchange based on task complexity.\\
\hline
\multirow{2}{*}{\cite{motlagh2025zkp, lavin2024zkp}} & \multirow{2}{*}{\shortstack[l]{Zero-Knowledge Proof \\ Protocols}} & (1) zk-SNARKs, zk-STARKs, Bulletproofs for distributed systems.\\
& & (2) Trade-offs in scalability, efficiency, and proof size for different environments.\\
\hline
\multirow{2}{*}{\cite{lavin2024zkp, lungu2025zkp, skorobogatova2023zkp}} & \multirow{2}{*}{\shortstack[l]{ZKP Applications in \\ Distributed Systems}} & (1) Blockchain: confidential transactions, private smart contracts.\\
& & (2) Cloud/IoT: privacy-preserving data sharing and model verification.\\
\hline
\multirow{2}{*}{\cite{li2023context, chen2022context}} & \multirow{2}{*}{\shortstack[l]{Context Compression \& \\ Efficiency}} & (1) Context compression techniques for large language models.\\
& & (2) Self-supervised representation learning for context understanding.\\
\hline 
\multirow{2}{*}{\cite{gao2024long, chen2023long}} & \multirow{2}{*}{\shortstack[l]{Long-Context \\ Processing}} & (1) Training long-context language models effectively.\\
& & (2) Efficient fine-tuning of long-context large language models.\\
\hline
\multirow{4}{*}{Our work} & \multirow{4}{*}{\shortstack[l]{ZK-based MCP \\ audit model}} & (1) Model Context Protocol provides standardized agent communication.\\
& & (2) Zero-knowledge proofs enable privacy-preserving audit verification.\\
& & (3) Capability persistence verification prevents agent degradation.\\
& & (4) Hash-based verification ensures communication integrity.\\
\hline
\end{tabular}
\end{table*}

\begin{table*}[!htb]
\centering
\caption{Related Works on MCP Security, Privacy, and Zero-Knowledge Proofs \label{tab:mcp_security_privacy}}
\small
\begin{tabular}{|p{3.5cm}|p{10cm}|}
\hline \textbf{Reference} & \textbf{Detailed descriptions}\\
\hline 
\cite{hou2025mcp, patil2025mcp, narajala2025mcp} & Implement MCP architecture, security frameworks, and enterprise-grade mitigation strategies for multi-agent systems without zero-knowledge audit verification.\\
\hline
\cite{krishnan2025mcp, vadlamani2025mcp, li2023context} & Solving the MCP-based agent communication coordination, adaptive protocols, and context efficiency problems with traditional methods.\\
\hline
\cite{motlagh2025zkp, lavin2024zkp, grilo2025zkp} & Develop zero-knowledge proof protocols and applications in distributed systems, blockchain, and IoT without MCP integration.\\
\hline
\multirow{3}{*}{Our work} & (1) Implement zero-knowledge proof-based audit framework for MCP protocol.\\
& (2) Prove that the framework provides privacy-preserving verification without exposing communication content.\\
& (3) Solve the capability persistence verification problem with streaming communication constraints.\\
\hline
\end{tabular}
\end{table*}


Our work addresses this gap by proposing the first zero-knowledge proof-based audit framework specifically designed for MCP-based agent communication systems. This framework combines MCP's context-aware communication capabilities with ZKP's privacy-preserving verification mechanisms, enabling verifiable audit trails without exposing sensitive communication content or compromising agent privacy. The detailed comparison of related works is presented in Tables~\ref{tab:mcp_comm_protocols} and~\ref{tab:mcp_security_privacy}.

\section{System Model and Problem Definition}\label{sec 3}\

\begin{figure*}[!t]
\centering
\includegraphics[width=0.85\textwidth]{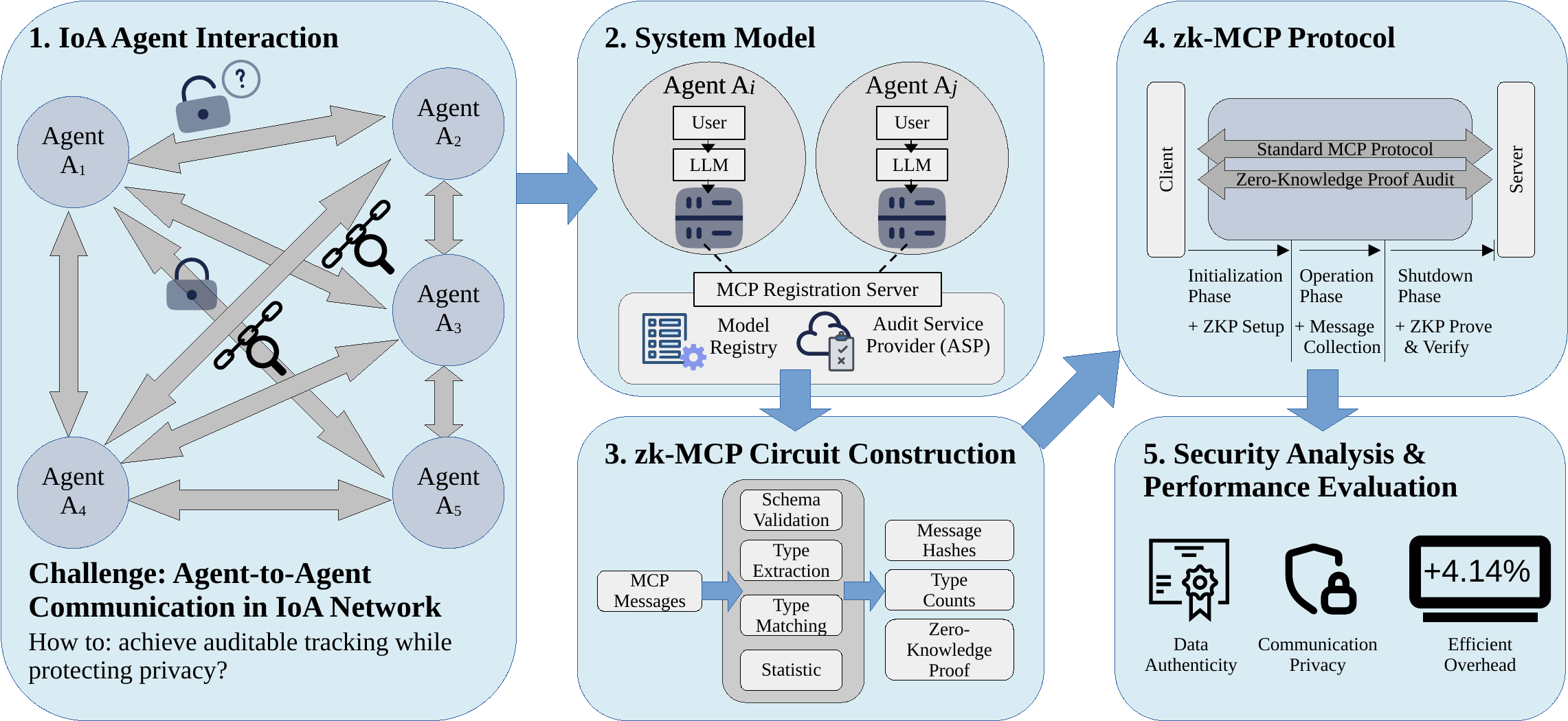}
\caption{Research Framework for MCP-based Privacy-Preserving Audit in Multi-Agent Communication using Zero-Knowledge Proofs. The framework addresses the challenge of auditing multi-agent communication in an IoA network while preserving privacy, leveraging zero-knowledge proofs. It details the system model, the zk-MCP protocol, the underlying ZKP circuit, and finally, the security analysis and performance evaluation.}
\label{fig:framework}
\end{figure*}

\subsection{System Model}

Our Zero-Knowledge Agent Communication Audit Framework involves five entities: agents (denoted as $\mathrm{A}_i$ where $i \in \{1, 2, \ldots, n\}$), MCP registry server, model registry (MR), and audit service provider (ASP). 

In this system, agents are autonomous entities that can participate in the IoA system. Each agent $\mathrm{A}_i$ has a unique identifier $\mathrm{ID}_{\mathrm{A}_i}$ and possesses a dual identity that enables it to act as both a service requester and a service provider simultaneously or at different times. This dual identity is fundamental to the IoA architecture, where agents form a complete peer-to-peer network where any agent can both request services from and provide services to other agents. Agents in IoA dynamically switch roles based on their current needs and capabilities, creating a fully decentralized and interconnected agent network.

Specifically, in any given interaction, an agent $\mathrm{A}_i$ can simultaneously or sequentially act as both a service requester and a service provider (e.g., $\mathrm{A}_1$ requests service from $\mathrm{A}_2$, while $\mathrm{A}_2$ simultaneously requests service from $\mathrm{A}_3$). $\mathrm{A}_i$ autonomously selects its service providers from available agents and initiates service requests through the MCP protocol, while also receiving requests from other agents and processing them using LLMs. $\mathrm{A}_i$ is responsible for generating three types of zero-knowledge proofs: (1) $\pi_{\mathrm{token}}^{\mathrm{A}_i}$ that demonstrates the actual number of tokens it has consumed (e.g., for request processing, verification, or other computational tasks) without revealing the specific content of its requests or internal computations; (2) $\pi_{\mathrm{output}}^{\mathrm{A}_i}$ that demonstrates the authenticity of the output content without revealing the input content, ensuring that the generated response is legitimate and corresponds to the input request. Each agent maintains a registered model configuration that includes the model type, version, and capability parameters. It is assumed that each agent is equipped with the MCP protocol stack to support bidirectional communication with other agents, and an agent's role can change dynamically across different interactions in the IoA ecosystem, forming a complete network of interconnected agents.

The MCP Registry Server  serves as the central communication hub in the IoA. The protocol follows a rigorous lifecycle consisting of initialization (capability negotiation and protocol version agreement), operation (normal protocol communication), and shutdown (graceful connection termination). Unlike centralized service matching systems, in our framework, each agent $\mathrm{A}_i$ autonomously selects its service providers from the available agents in the IoA ecosystem. MCP does not perform capability matching or service provider selection on behalf of agents; instead, it provides the communication infrastructure that enables agents to discover, connect, and communicate with each other through bidirectional message exchange (requests, responses, and notifications). 

All registered models form a group $\mathcal{M}$ managed by the Model Registry (MR). The MR maintains a comprehensive database of certified LLM models, each associated with a unique model identifier $\mathrm{ModelID}$, capability parameters (e.g., model size, training data, performance metrics), and a model commitment $\mathrm{cmt}_{\mathrm{Model}}$ that can be used for zero-knowledge verification. The MR's main task is to generate model authenticity proofs and support verification of whether an agent is using the claimed model quality. There can be thousands of registered models in the MR, providing diverse service capabilities for different agent requirements. The MR operates independently and is accessible to both agents and audit service providers for verification purposes.

The Audit Service Provider (ASP) is an independent, third-party certification and auditing organization that operates similarly to international standards organizations or certification bodies in regulated industries. The ASP serves as a neutral verification authority. The ASP has no stake in any transaction between agents. In enterprise and regulated environments where agents handle sensitive data—such as financial transactions, medical records, or proprietary business intelligence—the ASP provides compliance auditing, quality assurance, and accountability services. 

The ASP is responsible for verifying the zero-knowledge proofs submitted by agents through two separate audit processes within the same audit department. Our framework uses three distinct types of zero-knowledge proofs tailored to IoA's audit requirements: (1) \textbf{Token consumption proof} ($\pi_{\mathrm{token}}^{\mathrm{A}_i}$): Agents prove their actual token consumption without revealing request content; (2) \textbf{Output authenticity proof} ($\pi_{\mathrm{output}}^{\mathrm{A}_i}$): Agents prove output legitimacy without exposing input content;  The first audit process verifies the token consumption proof $\pi_{\mathrm{token}}^{\mathrm{A}_i}$ submitted by agents to ensure accurate billing based on the actual token usage. The second audit process verifies the proofs submitted by agents, including $\pi_{\mathrm{output}}^{\mathrm{A}_i}$ to ensure output content authenticity. The verification process is executed through an audit protocol that determines the legitimacy of the uploaded proofs. The ASP operates transparently and maintains comprehensive audit logs for accountability and compliance purposes, but does not participate in the billing or payment processing. In scenarios where multiple agents interact simultaneously (e.g., $\mathrm{A}_1$ requesting service from $\mathrm{A}_2$, while $\mathrm{A}_2$ simultaneously requests service from $\mathrm{A}_3$), the ASP can handle multiple concurrent audit processes independently, ensuring that each interaction is verified separately while maintaining the privacy and integrity of each audit trail.

The communication flow in our system proceeds as follows:
1) An agent $\mathrm{A}_i$ initiates a service request through MCP, specifying the required model capabilities and service parameters. MCP then facilitates the establishment of bidirectional communication channels between $\mathrm{A}_i$ and the selected agents, serving purely as a communication infrastructure without performing service matching or selection.
\\
2) The agents exchange messages bidirectionally through MCP, with agents processing requests using their LLMs. The MCP protocol maintains the communication channels and provides streaming support for continuous message exchange. Note that in complex scenarios, multiple agents can interact simultaneously (e.g., $\mathrm{A}_1$ requesting service from $\mathrm{A}_2$, while $\mathrm{A}_2$ simultaneously requests service from $\mathrm{A}_3$), creating a multi-agent interaction network.
\\
3) After service completion, each agent $\mathrm{A}_i$ generates three types of zero-knowledge proofs: (a) $\pi_{\mathrm{token}}^{\mathrm{A}_i}$ that demonstrates its actual token consumption $\mathrm{TokenCount}_{\mathrm{A}_i}$ during the service interaction (e.g., for request processing, verification, or other computational tasks) without revealing the specific content of its requests or internal computations; (b) $\pi_{\mathrm{output}}^{\mathrm{A}_i}$ that demonstrates the authenticity of the output content without revealing the input content, ensuring that the generated response is legitimate and corresponds to the input request. Each agent submits $\pi_{\mathrm{token}}^{\mathrm{A}_i}$ to the ASP for the first audit process, and submits $\pi_{\mathrm{output}}^{\mathrm{A}_i}$ to the ASP for the second audit process.
\\
4) The ASP performs two independent audit processes for each interaction. In the first audit process, the ASP verifies $\pi_{\mathrm{token}}^{\mathrm{A}_i}$ to ensure accurate billing based on $\mathrm{A}_i$'s actual token usage. In the second audit process, the ASP verifies $\pi_{\mathrm{output}}^{\mathrm{A}_i}$  to ensure output content authenticity and model quality compliance. As an independent third party, the ASP maintains audit logs and provides verification results to all participating agents without participating in billing or payment processing.

\section{zk-MCP: Privacy-Preserving MCP Audit Verification}
\label{sec:zk-mcp-protocol}
In this section, we address the audit verification problem for MCP-based agent communication in our IoA framework. We present zk-MCP, which uses zero-knowledge proofs to enable privacy-preserving audit verification for MCP communication. The core focus is on solving the MCP audit verification problem, where zero-knowledge proofs provide a natural fit: MCP communication requires audit verification to ensure statistical correctness, yet message content must remain private. Zero-knowledge proofs allow agents to prove the correctness of communication statistics without revealing the actual message content, making them an ideal solution for MCP audit verification. zk-MCP is proven to satisfy data authenticity and communication privacy in addition to standard zk-SNARK properties.

\subsection{ Overview}

In our IoA framework, we aim to enable an agent to securely prove its communication statistics and message type distribution without revealing the actual content of messages exchanged through MCP. However, under the assumption that agents can be untrusted, how to ensure the correctness and privacy of audit information becomes challenging.

In the following, we give the definition of zk-MCP.

\textbf{Definition 1 (zk-MCP).} zk-MCP is a zk-SNARK for arithmetic circuit, and is composed of 3 polynomial-time algorithms $\Delta_{\text{MCP}} \stackrel{\text{def}}{=}$ (Setup, Prove, Verify).

The construction of zk-MCP, $\Delta_{\text{MCP}}$, is based on a zk-SNARK algorithm $\Pi$ and works as follows:

- $(crs) \leftarrow \text{Setup}(1^{\lambda}, C)$. On inputs of a security parameter $\lambda$ and a circuit $C$, the algorithm runs $\Pi.\text{Setup}(1^{\lambda}, C)$ to produce a common reference string $crs$, and finally publishes $crs$.

\begin{figure}[h]
\centering
\includegraphics[width=0.44\textwidth]{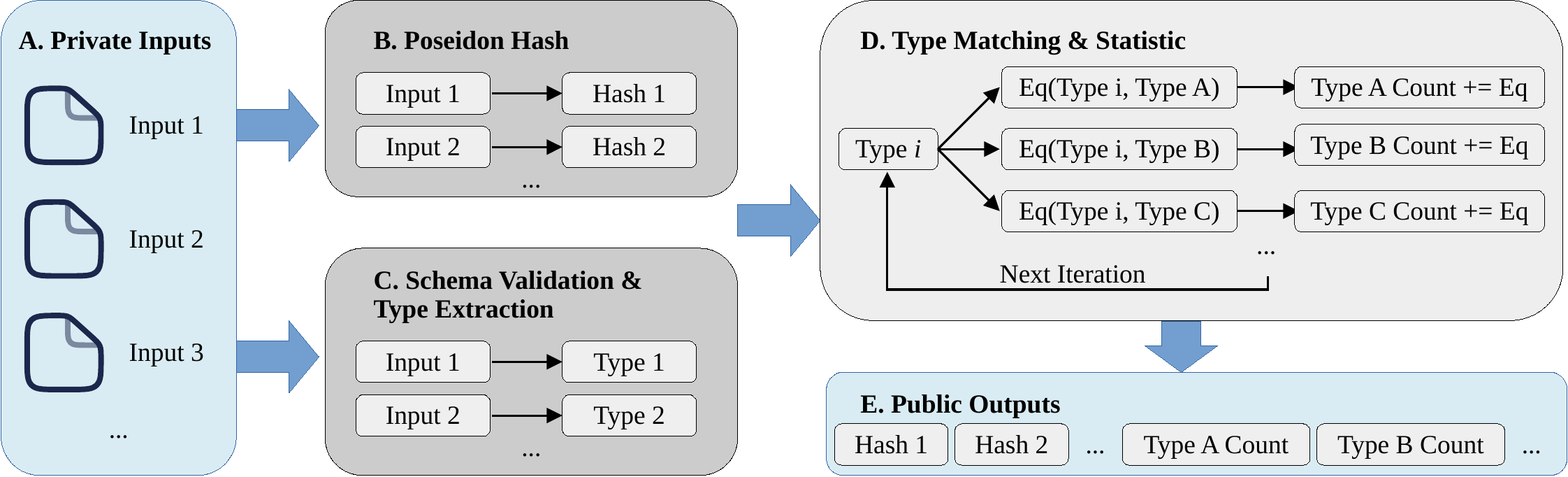}
\caption{The logic diagram of the zk-MCP circuit. }
\label{fig:zk_mcp_circuit}
\end{figure}

Note that $crs$ is related to the structure of $C$ (like the number of inputs and their operations) and is independent of the specific input assignment. The specific circuit logic used in zk-MCP is shown in Fig.~\ref{fig:zk_mcp_circuit}. The private input $w$ includes MCP message arrays $\text{json}[n][L]$ and type length arrays $\text{type\_len}[n]$, where $n$ is the number of messages and $L$ is the maximum JSON length. The public output includes type counts $\text{counts}[K]$ for $K$ different message types and message hashes $\text{hashes}[n]$ for verification purposes.

- $\pi \leftarrow \text{Prove}(crs, x, w)$. This algorithm generates a proof $\pi$ based on $\Pi.\text{Prove}(crs, x, w)$. $x$ represents the public values including type counts and hashes. $w$ represents the private values including messages content and type lengths. Users can generate their own proofs with the same $crs$ and different input assignments.

- $\{0,1\} \leftarrow \text{Verify}(crs, \pi, x)$. This algorithm runs $\Pi.\text{Verify}(crs, \pi, x)$ to verify the proof $\pi$ with the public values $x$ and the common reference string $crs$. It outputs 1 if the proof $\pi$ is valid and 0 otherwise.

\subsection{ Circuit Construction}

The zk-MCP circuit is constructed using Circom, a domain-specific language for zero-knowledge circuit design. The circuit template processes $n$ JSON messages and verifies their format while counting message types without revealing the actual message content.

\begin{table}[h]
\centering
\caption{Circuit Parameters, Inputs, and Outputs}
\label{tab:circuit_params}
\begin{minipage}{0.32\textwidth}
\centering
\begin{tabular}{p{0.4\linewidth}p{0.55\linewidth}}
\hline
\multicolumn{2}{c}{\textbf{Parameters}} \\
\hline
$n$ & Number of JSON messages \\
$\text{MAX\_JSON}$ & $64$ (max JSON length) \\
$\text{MAX\_TYPE}$ & $20$ (max type length) \\
$\text{NUM\_TYPES}$ & $8$  \\
\hline
\end{tabular}
\end{minipage}
\hfill
\begin{minipage}{0.32\textwidth}
\centering
\begin{tabular}{p{0.4\linewidth}p{0.55\linewidth}}
\hline
\multicolumn{2}{c}{\textbf{Private Inputs}} \\
\hline
$\text{json}[n][\text{MAX\_JSON}]$ &  JSON message bytes \\
$\text{type\_len}[n]$ & Type string lengths \\

\hline
\end{tabular}
\end{minipage}
\hfill
\begin{minipage}{0.32\textwidth}
\centering
\begin{tabular}{p{0.4\linewidth}p{0.55\linewidth}}
\hline
\multicolumn{2}{c}{\textbf{Public Outputs}} \\
\hline
$\text{counts}[\text{NUM\_TYPES}]$ & Count of each message type \\
$\text{hashes}[n]$ & Hash of each message \\

\hline
\end{tabular}
\end{minipage}
\end{table}

\textbf{Circuit Logic:}

The circuit enforces the following constraints for each JSON message $i \in [0, n-1]$:

 \textbf{Format Validation:} The circuit verifies that each JSON message follows the format $\{\text{``type'': ``X''}\}$ where X is a character type identifier. 

\textbf{Type Extraction:} The circuit extracts the type character of the JSON string, where the type length is determined by $\text{type\_len}[i]$. The extracted type bytes are padded with zeros to $\text{MAX\_TYPE}$ length.

\textbf{Type Matching:} The circuit compares the extracted type against known types and ensures exactly one match per message. This is enforced by:
   $$\sum_{j=0}^{\text{NUM\_TYPES}-1} \text{match}[i][j] = 1$$
   where $\text{match}[i][j]$ indicates whether message $i$ matches type $j$.

\textbf{Count Accumulation:} The circuit accumulates counts for each type:
   $$\text{sum}[j][i+1] = \text{sum}[j][i] + \text{match}[i][j]$$
   for each type $j \in [0, \text{NUM\_TYPES}-1]$.

\textbf{Hash Computation:} The circuit computes a Poseidon hash~\cite{grassi2021poseidon} of each JSON message for verification purposes. The hash input includes the packed JSON bytes and the total length, providing domain separation:
   $$\text{hashes}[i] = \text{Poseidon}(\text{json\_num}[i], \text{total\_len}[i])$$
   where $\text{json\_num}[i]$ is the packed representation of the JSON bytes and $\text{total\_len}[i]$ is the total JSON length.

\textbf{Padding Constraint:} The circuit enforces that bytes beyond the actual JSON length must be zero:
   $$\text{json}[i][m] \cdot \text{ge}[i][m].\text{out} = 0$$
   for all $m \geq \text{total\_len}[i]$, where $\text{ge}[i][m]$ is a greater-or-equal comparator.

The complete circuit algorithm is described in Algorithm~\ref{alg:zk_mcp_circuit}.

\begin{algorithm}
\caption{zk-MCP Circuit Algorithm\label{alg:zk_mcp_circuit}}
\nl $\text{json}[n][L]$: JSON message array; $\text{type\_len}[n]$: type length array; \\
\nl $\text{counts}[K]$: type counts; $\text{hashes}[n]$: message hashes; \\
\nl $\text{sum}[K]$: type sum array; $\text{type}[n]$: extracted type array; \\
\nl $\text{match}[n][K]$: type matching array; \\
\vspace{-5pt}
\hrulefill\\
\nl $\text{sum}[K] \gets 0$; \\
\nl \For{$i$ = $0$ to $n-1$}{ 
\nl \textbf{Step 1: Validate JSON format}; \\
\nl Verify: ``type'' exists in $\text{json}[i]$; \\
\nl Verify: $0$ $\leq$ $\text{json}[i][k]$ $\leq$ $255$ for MAX\_JSON; \\
\nl \textbf{Step 2: Extract and verify type}; \\
\nl $\text{type}[i] \gets \text{extract\_type}(\text{json}[i], \text{type\_len}[i])$; \\
\nl \textbf{Step 3: Type matching}; \\
\nl \For{$j$ = $0$ to $K-1$}{
\nl     $\text{match}[i][j] \gets (\text{type}[i]$ = $\text{known\_types}[j])$; }
\nl Verify: $\sum_{j=0}^{K-1} \text{match}[i][j]$ = $1$; \\
\nl \textbf{Step 4: Update statistics}; \\
\nl \For{$j$ = $0$ to $K-1$}{
\nl     $\text{sum}[j] \gets \text{sum}[j] + \text{match}[i][j]$; }
\nl \textbf{Step 5: Compute hash}; \\
\nl $\text{hashes}[i] \gets \text{Poseidon}(\text{json}[i], \text{type\_len}[i])$; }
\nl \textbf{Output results:}; \\
\nl \For{$j$ = $0$ to $K-1$}{
\nl     $\text{counts}[j] \gets \text{sum}[j]$; }
\hrulefill\\
\end{algorithm}

\subsection{Security Analysis}

The security properties of zk-MCP are defined as follows:

\textbf{Definition 1 (Data Authenticity).} The prover can convince the verifier that the output counts are computed from the private input JSON messages that match the public hashes, without revealing the actual message content. Formally, we define the following experiment:

\begin{equation*}
\mathbf{Exp}_{\text{zk-MCP}, \mathcal{A}}^{\text{auth}}(1^{\lambda}, C):
\end{equation*}
\begin{itemize}
    \item $(crs) \leftarrow \text{Setup}(1^{\lambda}, C)$
    \item $(\pi, x) \leftarrow \mathcal{A}(crs)$
    \item $w \leftarrow \mathcal{E}(\text{trans}_{\mathcal{A}})$
    \item If $((x, w) \notin \mathcal{R}_C$ and $\text{Verify}(crs, \pi, x)$ = $1)$ return 1
    \item Else return 0
\end{itemize}

Here, $\mathcal{A}$ is a non-uniform polynomial-time adversary, $\mathcal{R}_C$ is the relation defined by the circuit $C$, $\text{trans}_{\mathcal{A}}$ is the list containing all $\mathcal{A}$'s inputs, outputs, and randomness, and $\mathcal{E}$ is a probabilistic polynomial-time witness extractor.

We say that a zk-MCP scheme achieves data authenticity if for any non-uniform polynomial-time adversary $\mathcal{A}$, there exists a probabilistic polynomial-time witness extractor $\mathcal{E}$ such that $\Pr\left[\mathbf{Exp}_{\text{zk-MCP}, \mathcal{A}}^{\text{auth}}(1^{\lambda}, C) = 1\right]$ is negligible.

\textbf{Theorem 1.} If $\Pi$ is a zk-SNARK scheme satisfying completeness, soundness, succinctness, and zero knowledge, then the above construction of zk-MCP satisfies data authenticity.

\textbf{Proof.} The soundness of our construction can be easily reduced to the soundness of $\Pi$: if an adversary can produce a valid proof for a false statement, then the adversary can break the soundness of $\Pi$. For data authenticity, we need to show that the probability $\Pr\left[\mathbf{Exp}_{\text{zk-MCP}, \mathcal{A}}^{\text{auth}}(1^{\lambda}, C) = 1\right]$ is negligible. Due to the soundness of our zk-MCP construction, if $\text{Verify}(crs, \pi, x) = 1$, then with overwhelming probability, there exists a witness $w$ such that $(x, w) \in \mathcal{R}_C$. The hash function ensures that the witness $w$ corresponds to messages that match the public hashes. Since the hash function is collision-resistant, the probability that an adversary can find a witness $w'$ such that $(x, w') \notin \mathcal{R}_C$ but the hashes match is negligible. $\square$

\textbf{Definition 2 (Communication Privacy).} This property is defined as the notion that an adversary cannot learn any information about the actual message content from the proof and public outputs. More formally, we define the following experiment:

\begin{equation*}
\mathbf{Exp}_{\text{zk-MCP}, \mathcal{A}}^{\text{priv}}(1^{\lambda}, C):
\end{equation*}
\begin{itemize}
    \item $(crs) \leftarrow \text{Setup}(1^{\lambda}, C)$
    \item $(m_0, m_1, \text{state}) \leftarrow \mathcal{A}_1^{\text{Setup}(\cdot),\ \text{Prove}(\cdot),\ \text{Verify}(\cdot)}(crs)$
    \item $b \leftarrow_R \{0,1\}$
    \item $\pi^* \leftarrow \text{Prove}(crs, x_b, w_b)$ \ \ \textit{where $x_b$ and $w_b$ correspond to message $m_b$}
    \item $b' \leftarrow \mathcal{A}_2^{\text{Setup}(\cdot),\ \text{Prove}(\cdot),\ \text{Verify}(\cdot)}(\pi^*,\, \text{state})$
    \item If $b'$ = $b$ return 1 else return 0
\end{itemize}

We say that a zk-MCP scheme achieves communication privacy if for any non-uniform polynomial-time adversary $\mathcal{A}$, the probability $\Pr\left[\mathbf{Exp}_{\text{zk-MCP}, \mathcal{A}}^{\text{priv}}(1^{\lambda}, C) = 1\right] - \frac{1}{2}$ is negligible.

\textbf{Theorem 2.} If $\Pi$ is a zk-SNARK scheme satisfying completeness, soundness, succinctness, and zero knowledge, then the above construction of zk-MCP satisfies communication privacy.

\textbf{Proof.} For communication privacy, the zero-knowledge property of $\Pi$ ensures that the proof $\pi$ does not reveal any information about the witness $w$ beyond what is implied by the public input $x$. Since the public input $x$ only contains type counts and message hashes, and the hash function is one-way, the adversary cannot learn the actual message content from the proof. The indistinguishability between proofs for different messages follows directly from the zero-knowledge property of $\Pi$. $\square$

\subsection{zk-MCP Protocol: A Privacy-Preserving MCP Audit Framework}

In this subsection, we present the zk-MCP protocol, which integrates zero-knowledge proofs with MCP to enable privacy-preserving audit verification for agent communication. Our protocol operates in a lightweight network environment where agents communicate through MCP and generate zero-knowledge proofs for audit purposes. The key difference from standard MCP is that zk-MCP enables verifiable audit capabilities without compromising communication efficiency or privacy.

The zk-MCP protocol extends standard MCP communication with zero-knowledge proof-based audit verification. During normal MCP communication between agents $\mathrm{A}_i$ and $\mathrm{A}_j$, messages are exchanged following the standard MCP lifecycle (Initialization, Operation, Shutdown). The protocol maintains full compatibility with standard MCP, ensuring that communication efficiency. After each communication session, agents can generate zero-knowledge proofs to demonstrate their communication statistics without revealing message content.

The protocol consists of four main phases: (1) \textbf{Initialize}: Agents establish MCP connections and obtain the common reference string $crs$; (2) \textbf{Communication}: Agents exchange messages through standard MCP protocol while collecting message metadata; (3) \textbf{Proof Generation}: Agents generate zero-knowledge proofs for communication statistics; (4) \textbf{Audit Verification}: The  ASP verifies proofs and records audit results.

The zk-MCP protocol involves three main entities:
\begin{itemize}
    \item \textbf{Agent $\mathrm{A}_i$ (Prover)}: The agent that generates zero-knowledge proofs for its communication statistics
    \item \textbf{Agent $\mathrm{A}_j$ (Communicating Party)}: The agent that communicates with $\mathrm{A}_i$ through standard MCP protocol
    \item \textbf{Audit Service Provider (ASP) (Verifier)}: The entity that verifies zero-knowledge proofs and maintains audit records
\end{itemize}

\begin{figure}[h]
\centering
\includegraphics[width=0.475\textwidth]{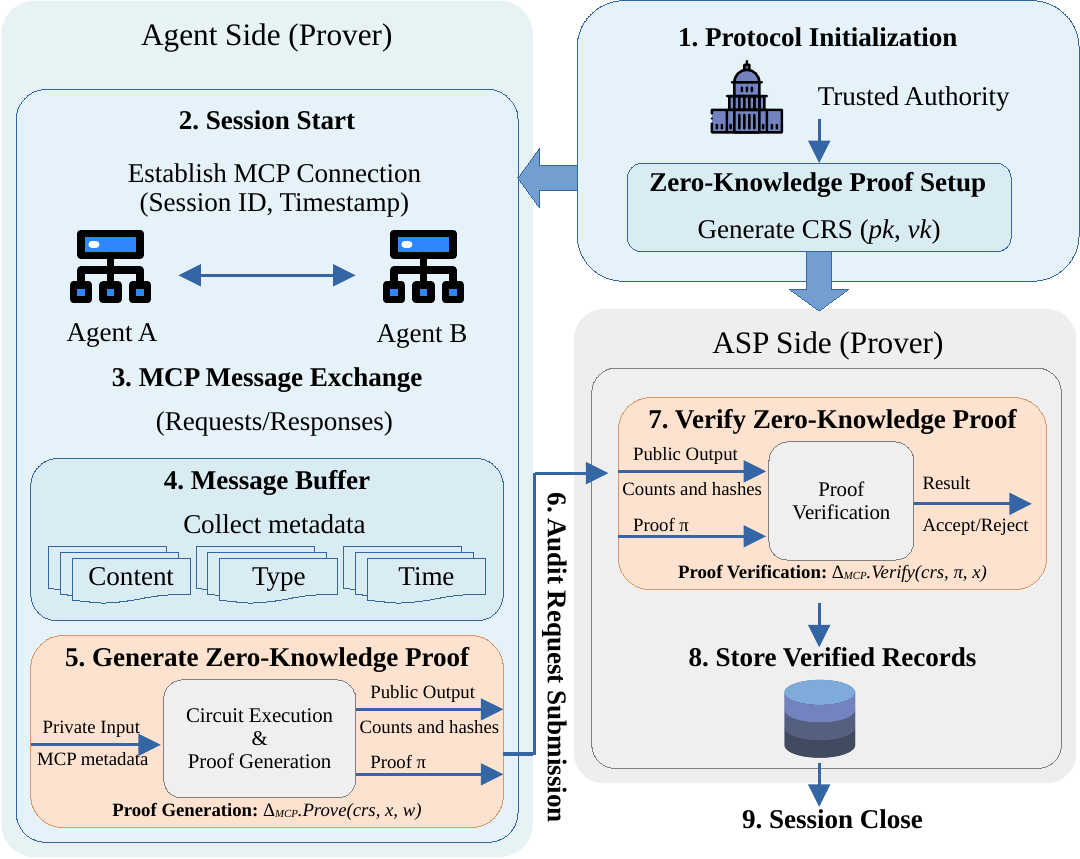}
\caption{Architecture of zk-MCP: Privacy-Preserving Auditing for Agent Communication. The diagram shows the complete workflow including protocol initialization, MCP message exchange, zero-knowledge proof generation on the Agent Side (Prover), and privacy-preserving verification on the ASP Side (Verifier).}
\label{fig:zk_mcp_architecture}
\end{figure}

Figure~\ref{fig:zk_mcp_architecture} illustrates the architecture of zk-MCP, showing the complete workflow from protocol initialization to session closure, including the interaction between agents and the ASP, message exchange, zero-knowledge proof generation, and audit verification.

Table~\ref{tab:zk_mcp_params} summarizes the key parameters used in the zk-MCP protocol. These parameters define the protocol's configuration and determine the security and efficiency properties of the system.

\begin{table}[h]
\centering
\caption{zk-MCP Protocol Parameters}
\label{tab:zk_mcp_params}
\begin{tabular}{|l|l|}
\hline
\textbf{Parameter} & \textbf{Description} \\
\hline
$n$ & Number of messages in a communication session \\
$L$ & Maximum JSON message length (bytes) \\
$K$ & Number of message types \\
$\lambda$ & Security parameter \\
$crs$ & Common reference string (proving + verification key) \\
$\pi$ & Zero-knowledge proof \\
$\text{counts}[K]$ & Public output: type counts \\
$\text{hashes}[n]$ & Public output: message hashes \\
$\text{json}[n][L]$ & Private input: message content \\
$\text{type\_len}[n]$ & Private input: type lengths \\
$\text{s\_id}$ & Unique identifier for each communication session \\
$\text{msg\_count}$ & Counter for messages in current session \\
\hline
\end{tabular}
\end{table}

The protocol parameters are defined as follows:
\begin{itemize}
    \item $n$: The number of messages exchanged in a single communication session between agents $\mathrm{A}_i$ and $\mathrm{A}_j$. This parameter determines the size of the zero-knowledge proof circuit.
    \item $L$: The maximum length of a JSON message in bytes. Messages are padded to this length to ensure uniform circuit structure.
    \item $K$: The number of distinct message types that can be identified in the protocol. 
    \item $\lambda$: The security parameter that determines the cryptographic strength of the zero-knowledge proof system. 
    \item $crs$: The common reference string generated during the setup phase. It consists of a proving key $pk$ and a verification key $vk$, where $crs$ = $(pk, vk)$.
    \item $\pi$: The zero-knowledge proof generated by agent $\mathrm{A}_i$ to prove the correctness of communication statistics without revealing message content.
    \item $\text{counts}[K]$: The public output array containing the count of each message type. This is computed from the private messages and made public for audit verification.
    \item $\text{hashes}[n]$: The public output array containing Poseidon hashes of each message. These hashes serve as commitments to the actual messages.
    \item $\text{json}[n][L]$: The private input array containing the actual message content. This remains hidden from the verifier.
    \item $\text{type\_len}[n]$: The private input array containing the length of the type string for each message.
\end{itemize}

We define the zk-MCP Protocol Specification as the core protocol logic that governs audit verification. The protocol maintains state for each agent communication session and processes audit requests as detailed below. The protocol state machine consists of six states: \texttt{INIT}, \texttt{SESSION\_ACTIVE}, \texttt{AUDIT\_PENDING}, \texttt{AUDIT\_VERIFIED}, \texttt{AUDIT\_REJECTED}, and \texttt{SESSION\_CLOSED}.

Protocol initialization begins by setting the state to \texttt{INIT}. The common reference string $crs$ is generated via $\Delta_{\text{MCP}}.\mathrm{Setup}(1^{\lambda}, C)$, where the setup algorithm outputs $crs = (pk, vk)$, with $pk$ as the proving key and $vk$ as the verification key. The $crs$ is then published to all agents and the ASP. The protocol also initializes an empty audit database $\mathrm{AuditDB}$ and an empty session registry $\mathrm{Sessions}$.

Upon receiving a (\texttt{Session-Start}, $A_i$, $A_j$, $\mathit{s\_id}$) message, the protocol asserts that the current state is either \texttt{INIT} or \texttt{SESSION\_CLOSED}. The state is then set to \texttt{SESSION\_ACTIVE}. Session metadata is recorded as $\mathrm{Session}[\mathit{s\_id}] = (A_i,\;\allowbreak A_j,\; \mathit{s\_id},\allowbreak\; \mathit{start\_time},\allowbreak\; \texttt{SESSION\_ACTIVE})$. The message counter is initialized to $\mathrm{msg\_count}[\mathit{s\_id}] = 0$, and an empty message buffer $\mathrm{Messages}[\mathit{s\_id}]$ is created. The session identifier is added via $\mathrm{Sessions}.\mathrm{append}(\mathit{s\_id})$.

During MCP communication, agents $\mathrm{A}_i$ and $\mathrm{A}_j$ exchange messages via MCP following the standard MCP lifecycle without any modification to the MCP protocol itself. For each message $m_k$ exchanged, the protocol parses the message to extract the type identifier as $t_k = \text{extract\_type}(m_k)$. Each agent locally stores the message information as $\text{Messages}[\text{s\_id}][k] = (m_k, \allowbreak t_k, \allowbreak\text{timestamp}_k)$ and increments the message counter $\text{msg\_count}[\text{s\_id}] = \text{msg\_count}[\text{s\_id}] + 1$. After $n$ messages are exchanged, we have $\text{msg\_count}[\text{s\_id}] = n$. Zero-knowledge proof generation is performed after MCP communication completes.

Upon receiving an (\texttt{Audit-Request}, $\pi$, $\text{counts}[K]$, $\text{hashes}[n]$, $\text{s\_id}$) message from $\mathrm{A}_i$, the protocol first asserts that the current state is either \texttt{SESSION\_ACTIVE} or \texttt{SESSION\_CLOSED}, then sets the state to \texttt{AUDIT\_PENDING}. The proof verification is performed based on zk-MCP $\Delta_{\text{MCP}}$ by setting the public input as $x = (\text{counts}[K], \allowbreak\text{hashes}[n])$ and executing $b \leftarrow \Delta_{\text{MCP}}.\text{Verify}(crs, \allowbreak\pi, \allowbreak x)$. The verification returns:
$$b = \begin{cases}
1 & \text{if } \Delta_{\text{MCP}}.\text{Verify}(crs, \pi, x) = 1 \\
0 & \text{otherwise}
\end{cases}$$
If $b = 1$, the protocol sets the state to \texttt{AUDIT\_VERIFIED}. It records the audit result as $\text{AuditDB}.\text{append}((\text{s\_id}, \allowbreak\text{counts}[K], \allowbreak\text{hashes}[n], \allowbreak\text{timestamp}))$, updates the session state to $\text{Session}[\text{s\_id}].\text{state} = \texttt{AUDIT\_VERIFIED}$, and accepts the audit statistics. Otherwise, if $b = 0$, the protocol sets the state to \texttt{AUDIT\_REJECTED}. It records a violation as $\text{Violations}.\text{append}((\text{s\_id}, \allowbreak\text{timestamp}))$, updates the session state to $\text{Session}[\text{s\_id}].\text{state} = \texttt{AUDIT\_REJECTED}$, flags a potential violation, and rejects the audit result.

Upon receiving a (\texttt{Session-Close}, $\text{s\_id}$) message, the protocol asserts that the current state is either \texttt{SESSION\_ACTIVE}, \texttt{AUDIT\_VERIFIED}, or \texttt{AUDIT\_REJECTED}. The state is then set to \texttt{SESSION\_CLOSED}, and the session end time is recorded as $\text{Session}[\text{s\_id}].\text{end\_time}$. The session state is updated to \texttt{SESSION\_CLOSED}.

\textbf{Multi-Session Communication Protocol}. The following protocol describes the complete zk-MCP communication flow for multiple communication sessions between agents, demonstrating how zero-knowledge proofs are integrated without affecting MCP communication efficiency. The protocol supports concurrent multiple sessions, where each session maintains its own state and generates independent zero-knowledge proofs.

Agent $\mathrm{A}_i$ obtains the common reference string $crs$ from a trusted authority and initializes the session state. The agent initializes an empty session registry $\text{ActiveSessions}$ and empty message buffers $\text{MessageBuffers}$.

For each communication session with agent $\mathrm{A}_j$, agent $\mathrm{A}_i$ generates a unique session identifier as $\text{s\_id} = \text{generate\_id}(\mathrm{A}_i,\allowbreak \mathrm{A}_j, \allowbreak\text{timestamp})$ and sends a (\texttt{Session-Start}, $\mathrm{A}_i$, $\mathrm{A}_j$, $\text{s\_id}$) message to the ASP. The agent then establishes an MCP connection with $\mathrm{A}_j$ following standard MCP initialization as described in Sec.~3.5. Session-specific variables are initialized. The session identifier is added via $\text{ActiveSessions}.\text{append}(\text{s\_id})$.

During the MCP operation phase, agent $\mathrm{A}_i$ exchanges messages with $\mathrm{A}_j$ through the standard MCP protocol. For each message $m_k$ sent or received ($k = 1, 2, \ldots, n$), the message is processed via MCP with normal MCP operation and no delay. In parallel, the agent parses $m_k$ to extract the type identifier, where the type is extracted as $t_k = \text{extract\_type}(m_k)$ and the type length is extracted as $\text{type\_len}_k = \text{length}(t_k)$. The message information is stored locally as $\text{Messages}[\text{s\_id}][k] = (m_k, \allowbreak t_k, \allowbreak\text{timestamp}_k)$, and the message counter is incremented as $\text{msg\_count}[\text{s\_id}] = \text{msg\_count}[\text{s\_id}] + 1$. After $n$ messages are exchanged, we have $\text{msg\_count}[\text{s\_id}] = n$. Message storage and parsing occur asynchronously in parallel with MCP communication, ensuring that MCP communication latency is not affected. The time complexity for message parsing is $O(1)$ per message, and storage operations are non-blocking.

After the MCP session completes during the standard MCP shutdown phase, the agent prepares private inputs from the collected messages. For each message $k = 1, 2, \ldots, n$, the message $m_k$ is converted to JSON bytes as $\text{json}[\text{s\_id}][k] = \text{to\_bytes}(m_k)$ and padded to length $L$ as $\text{json}[\text{s\_id}][k][L] = \text{pad}(\text{json}[\text{s\_id}][k], L)$. The private input is then $w = (\text{json}[\text{s\_id}][n][L], \allowbreak\text{type\_len}[\text{s\_id}][n])$. The agent computes public inputs by computing type counts for each type $j = 0, 1, \ldots, K-1$:
$$\text{counts}[\text{s\_id}][j] = \sum_{k=1}^{n} \mathbf{1}[t_k = \text{type}_j]$$
where $\mathbf{1}[t_k = \text{type}_j]$ is the indicator function that equals 1 if $t_k = \text{type}_j$ and 0 otherwise. For each message $k = 1, 2, \ldots, n$, message hashes are computed as:
$$\text{hashes}[\text{s\_id}][k] = \text{Poseidon}(\text{json\_num}[\text{s\_id}][k], \text{total\_len}[\text{s\_id}][k])$$. The public input is $x = (\text{counts}[\text{s\_id}][K], \allowbreak\text{hashes}[\text{s\_id}][n])$. The agent then generates a zero-knowledge proof as $\pi[\text{s\_id}] \leftarrow \Delta_{\text{MCP}}.\text{Prove}(crs, \allowbreak x, \allowbreak w)$ and sends an (\texttt{Audit-Request}, $\pi[\text{s\_id}]$, $\text{counts}[\text{s\_id}][K]$, $\text{hashes}[\text{s\_id}][n]$, $\text{s\_id}$) to the ASP.

The agent closes the MCP connection during the standard MCP shutdown phase and records the end time as $\text{end\_time}[\text{s\_id}] = \text{current\_time}()$. The session duration is calculated as $\text{duration}[\text{s\_id}] = \text{end\_time}[\text{s\_id}] - \text{start\_time}[\text{s\_id}]$. The agent sends a (\texttt{Session-Close}, $\text{s\_id}$) message to the ASP, removes the session via $\text{ActiveSessions}.\text{remove}(\text{s\_id})$, and clears the message buffer by setting $\text{Messages}[\text{s\_id}] = \text{null}$.

The ASP obtains the common reference string $crs$ from a trusted authority and initializes the audit state. The ASP initializes an empty audit database $\text{AuditDB}$, an empty violation log $\text{Violations}$, and an empty session registry $\text{SessionRegistry}$ to track all communication sessions and their audit status. Upon receiving a (\texttt{Session-Start}, $\mathrm{A}_i$, $\mathrm{A}_j$, $\text{s\_id}$) message, the ASP records the session metadata as $\text{SessionRegistry}[\text{s\_id}] = (\mathrm{A}_i, \allowbreak\mathrm{A}_j, \allowbreak\text{start\_time}, \allowbreak\text{state} = \texttt{SESSION\_ACTIVE})$ and sets $\text{AuditState}[\text{s\_id}] = \texttt{PENDING}$.

Upon receiving an (\texttt{Audit-Request}, $\pi$, $\text{counts}[K]$, $\text{hashes}[n]$, $\text{s\_id}$) message, the ASP sets the public input as $x = (\text{counts}[K], \allowbreak\text{hashes}[n])$ and verifies the zero-knowledge proof by executing $b \leftarrow \Delta_{\text{MCP}}.\text{Verify}(crs, \allowbreak\pi, \allowbreak x)$. If $b = 1$, the ASP accepts the audit result, records the audit statistics as $\text{AuditDB}.\text{append}((\text{s\_id}, \allowbreak\text{counts}[K], \allowbreak\text{hashes}[n], \allowbreak\text{timestamp}))$, updates the session audit state to $\text{AuditState}[\text{s\_id}] = \texttt{VERIFIED}$, and updates the session registry to $\text{SessionRegistry}[\text{s\_id}].\text{state} = \texttt{AUDIT\_VERIFIED}$. Otherwise, if $b = 0$, the ASP rejects the audit result, records a violation as $\text{Violations}.\text{append}((\text{s\_id}, \allowbreak\text{timestamp}, \allowbreak\text{reason} = \text{``Invalid proof''}))$, updates the session audit state to $\text{AuditState}[\text{s\_id}] = \texttt{REJECTED}$, updates the session registry to $\text{SessionRegistry}[\text{s\_id}].\text{state} = \texttt{AUDIT\_REJECTED}$, and flags a potential violation.

Upon receiving a (\texttt{Session-Close}, $\text{s\_id}$) message, the ASP records the session end time as $\text{current\_time}()$, updates the session state to \texttt{SESSION\_CLOSED}, and calculates session statistics as $\text{SessionStats}[\text{s\_id}] = (\text{duration}, \allowbreak\text{msg\_count}, \allowbreak\text{audit\_status})$.

The complete multi-session protocol is described in Algorithm~\ref{alg:zk_mcp_protocol}.

\begin{algorithm}[!t]
\caption{zk-MCP Multi-Session Protocol\label{alg:zk_mcp_protocol}}
\nl $crs$: common reference string; $\pi$: zero-knowledge proof; \\
\nl $\text{Messages}$: message buffer; $\text{msg\_count}$: message counter; \\
\nl $\text{s\_id}$: session identifier; $\text{state}$: protocol state; \\
\nl $w$: private input; $x$: public input; $b$: verification result; \\
\vspace{-5pt}
\hrulefill\\
\nl \textbf{Setup Phase:}; \\
\nl $(crs) \leftarrow \Delta_{\text{MCP}}.\text{Setup}(1^{\lambda}, C)$; \\
\nl Publish $crs$ to all agents and ASP; \\
\vspace{-5pt}
\hrulefill\\
\nl \textbf{Agent $\mathrm{A}_i$ Protocol:}; \\
\nl \textbf{Initialize:}; \\
\nl Get $crs$ from trusted authority; \\
\nl $\text{state} \gets \texttt{INIT}$; \\
\nl \textbf{For each communication session:}; \\
\nl \textbf{Session Start:}; \\
\nl Send (\texttt{Session-Start}, $\mathrm{A}_i$, $\mathrm{A}_j$, $\text{s\_id}$) to ASP; \\
\nl Establish MCP connection with $\mathrm{A}_j$; \\
\nl $\text{Messages} \gets []$; $\text{msg\_count} \gets 0$; \\
\nl \textbf{Communication Phase:}; \\
\nl \For{$k = 1$ to $n$}{
\nl     Send/receive message $m_k$ via MCP; \\
\nl     Parse $m_k$ to extract type $t_k$ in parallel; \\
\nl     $\text{Messages}[k] \gets (m_k, t_k, \text{timestamp}_k)$; \\
\nl     $\text{msg\_count} \gets \text{msg\_count} + 1$; }
\nl \textbf{Proof Generation (After MCP Session):}; \\
\nl Close MCP connection; \\
\nl Prepare $w = (\text{json}[n][L], \text{type\_len}[n])$ from Messages; \\
\nl Compute $x = (\text{counts}[K], \text{hashes}[n])$ from Messages; \\
\nl $\pi \leftarrow \Delta_{\text{MCP}}.\text{Prove}(crs, x, w)$; \\
\nl Send (\texttt{Audit-Request}, $\pi$, $\text{counts}[K]$, $\text{hashes}[n]$, $\text{s\_id}$) to ASP; \\
\nl Send (\texttt{Session-Close}, $\text{s\_id}$) to ASP; \\
\vspace{-5pt}
\hrulefill\\
\nl \textbf{ASP Protocol:}; \\
\nl \textbf{Initialize:}; \\
\nl Get $crs$ from trusted authority; \\
\nl $\text{state} \gets \texttt{INIT}$; \\
\nl \textbf{Audit Verification:}; \\
\nl Upon receiving (\texttt{Audit-Request}, $\pi$, $\text{counts}[K]$, $\text{hashes}[n]$, $\text{s\_id}$): \\
\nl $x \gets (\text{counts}[K], \text{hashes}[n])$; \\
\nl $b \leftarrow \Delta_{\text{MCP}}.\text{Verify}(crs, \pi, x)$; \\
\nl \If{$b = 1$}{
\nl     Accept audit result; \\
\nl     Record: $(\text{s\_id}, \text{counts}[K], \text{timestamp})$; }
\nl \Else{
\nl     Reject audit result; \\
\nl     Flag violation: $(\text{s\_id}, \text{timestamp})$; }
\hrulefill\\
\end{algorithm}





The zk-MCP protocol provides the following security and privacy guarantees. The actual content of messages exchanged between $\mathrm{A}_i$ and $\mathrm{A}_j$ remains private. The ASP only receives type counts and message hashes, which do not reveal the actual message content due to the one-way property of hash functions and the zero-knowledge property of the proof. The ASP can verify that the reported type counts are correctly computed from the actual messages without accessing the message content, which is guaranteed by the soundness property of the zk-SNARK scheme. The protocol maintains the efficiency of standard MCP communication while adding audit capabilities, as proof generation and verification are performed asynchronously, ensuring that communication latency is not affected.

%
%

\section{Performance Evaluation}
\label{sec:performance-experiment}
In this section, we evaluate the performance of zk-MCP through comprehensive experiments. Our evaluation focuses on two key aspects: (1) the scalability of zero-knowledge proof generation with respect to circuit parameters, and (2) the impact of zk-MCP on standard MCP communication performance. The experimental results demonstrate that zk-MCP achieves privacy-preserving audit verification, with proof generation overhead remaining below 4.14\% of total communication costs due to asynchronous processing design.

\subsection{Experiment Setup}

Our experimental evaluation is conducted on two machines equipped with AMD 6850H processors (8 cores @ 3.2GHz) and 32GB RAM, running Arch Linux with kernel version 5.15.62.1. The MCP servers and zero-knowledge proof components are deployed on these machines to simulate agents communication environment.

The zk-MCP circuit is implemented using Circom, a domain-specific language for zero-knowledge circuit design. The zero-knowledge proof framework is built on snarkjs, which provides the proving and verification algorithms for zk-SNARKs. The experimental MCP tools are implemented using the Python-based fastmcp library, which enables efficient MCP protocol implementation. The circuit parameters are configured as follows: $\text{MAX\_JSON} = 64$ bytes, $\text{MAX\_TYPE} = 20$ bytes, and $\text{NUM\_TYPES} = 8$, as specified in Table~\ref{tab:circuit_params}.

The experimental setup includes three LLM models for agent communication: DeepSeek V3, GPT-4.1mini, and GPT-3.5 turbo. Each agent communicates through standard MCP protocol, and zero-knowledge proofs are generated asynchronously after communication sessions complete, following the zk-MCP protocol specification described in Sec.~4.4.

\subsection{Experiment 1: Circuit Parameter Scalability}

The first experiment evaluates the scalability of zk-MCP with respect to the number of messages processed by the zero-knowledge proof circuit. We vary the number of MCP messages $n$ from $2^0$ to $2^9$ and measure the following performance metrics: (1) CPU time consumption for Setup, Prove, and Verify phases; (2) memory consumption during proof generation and verification; (3) circuit size metrics including constraint count and gate count; and (4) proof size and system parameter size (verification key and proving key) in bytes.

The circuit construction follows the zk-MCP specification in Sec.~4.2, where each message undergoes format validation, type extraction, type matching, count accumulation, and hash computation. As $n$ increases, the circuit must process more messages, leading to increased computational complexity and circuit size.

\begin{figure}[!t]
\centering
\includegraphics[width=0.48\textwidth]{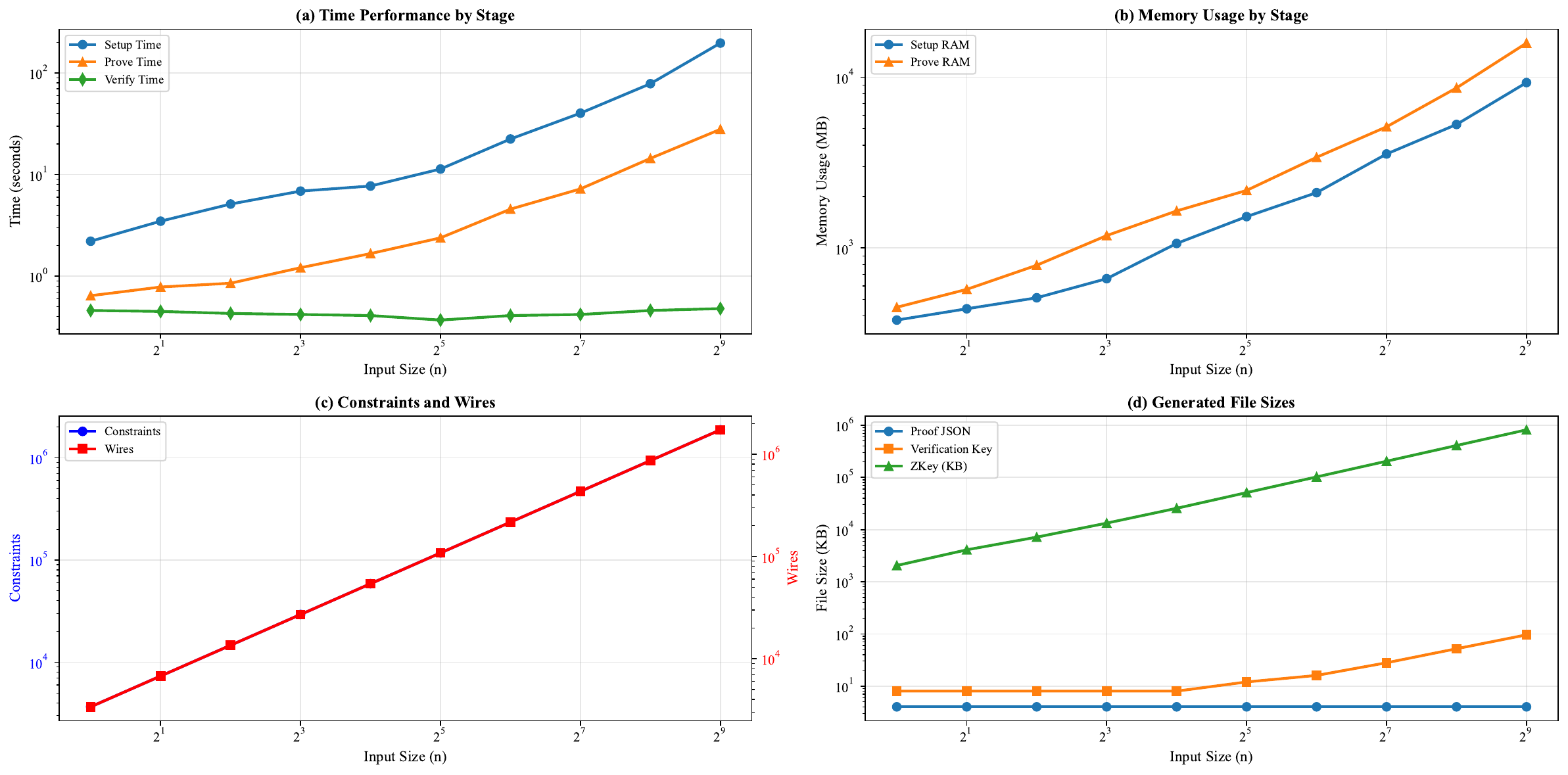}
\caption{Performance metrics of zk-MCP circuit with varying input size $n$: (a) Time performance comparison across Setup, Prove, and Verify phases; (b) Memory usage during Setup and Prove phases; (c) Circuit complexity metrics (constraints and wires); (d) Generated file sizes including proof size and system parameter size (verification key and proving key). All metrics are plotted on logarithmic scales to accommodate the wide range of values.}
\label{fig:circuit_scalability}
\end{figure}

Figure~\ref{fig:circuit_scalability} presents the experimental results. The results demonstrate that as the number of messages $n$ increases, all performance metrics show a corresponding increase. Specifically, as shown in Fig.~\ref{fig:circuit_scalability}(a), the Setup phase time grows super-linearly with $n$, as the common reference string generation depends on the circuit structure. The Prove phase time exhibits a super-linear growth pattern, reflecting the increased complexity of generating proofs for larger message sets. The Verify phase remains relatively constant, as verification time is primarily determined by proof size rather than the number of messages.

As illustrated in Fig.~\ref{fig:circuit_scalability}(b), memory consumption during Setup and Prove phases increases with $n$, with Prove phase requiring significantly more memory than Setup phase. The circuit constraint count and wire count, shown in Fig.~\ref{fig:circuit_scalability}(c), both increase proportionally with $n$, as each additional message requires additional constraints for format validation, type matching, and hash computation. The proof size and system parameter size (verification key and proving key), presented in Fig.~\ref{fig:circuit_scalability}(d), also increase with $n$, though the growth rate is sub-linear due to the succinctness property of zk-SNARKs.

\begin{figure}[!t]
\centering
\includegraphics[width=0.48\textwidth]{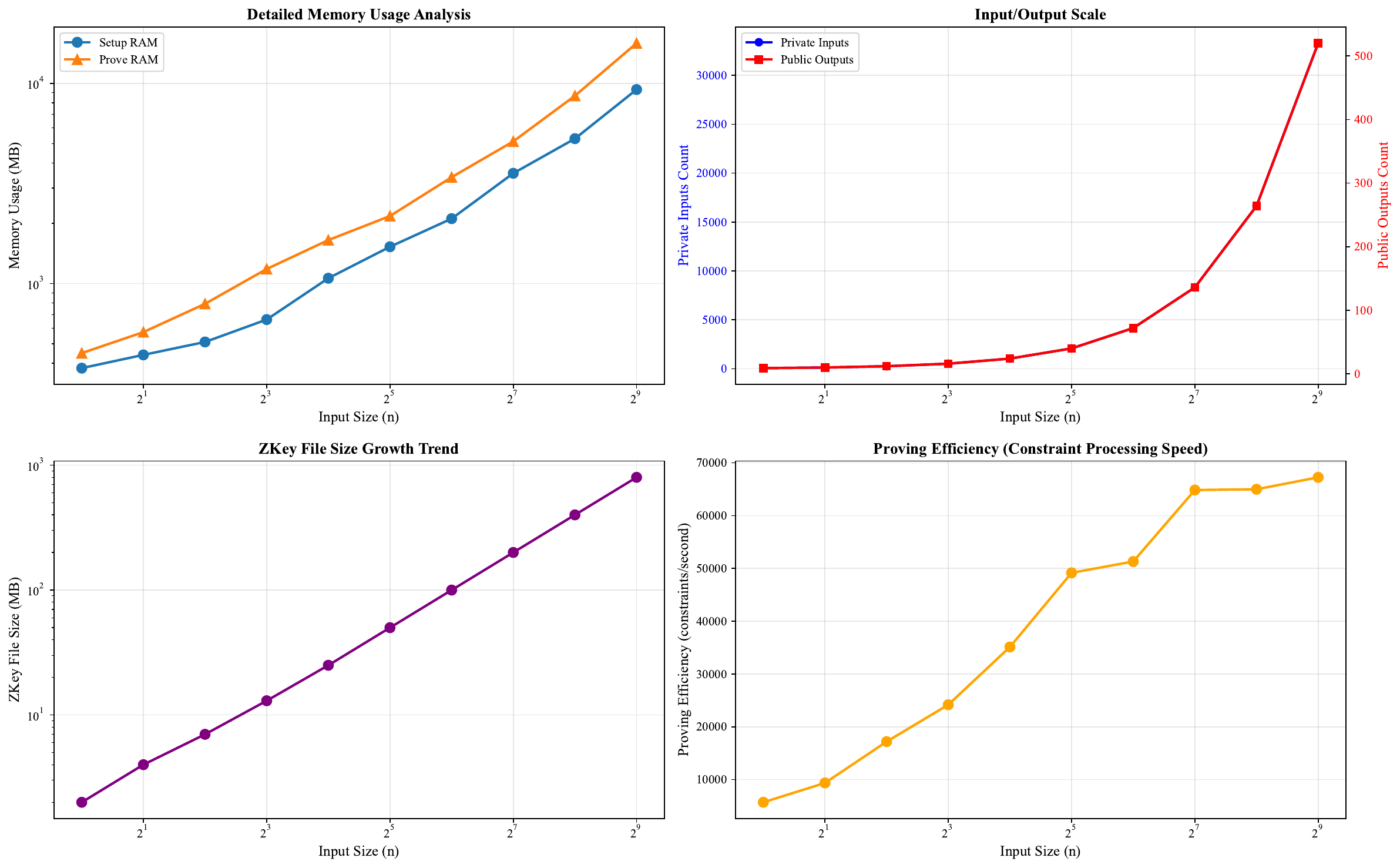}
\caption{Detailed performance analysis of zk-MCP circuit: (a) Detailed memory usage analysis for Setup and Prove phases; (b) Input/output scale showing private inputs and public outputs count; (c) Proving key file size growth trend; (d) The average number of constraints processed per second}
\label{fig:detailed_analysis}
\end{figure}

Figure~\ref{fig:detailed_analysis} provides additional detailed analysis of the circuit performance. Fig.~\ref{fig:detailed_analysis}(a) shows the detailed memory usage patterns, confirming that memory consumption scales with circuit size. Fig.~\ref{fig:detailed_analysis}(b) illustrates the input/output scale, demonstrating how private inputs and public outputs scale with $n$. The proving key file size growth trend in Fig.~\ref{fig:detailed_analysis}(c), together with proof size and system parameter size shown in Fig.~\ref{fig:circuit_scalability}(d), demonstrates efficient storage and communication overhead. Finally, Fig.~\ref{fig:detailed_analysis}(d) presents the proving efficiency, measured as the number of constraints processed per second, which remains relatively stable across different values of $n$, indicating consistent proving performance.

These results demonstrate the feasibility of using zero-knowledge proofs for audit verification, as the circuit successfully processes messages of varying counts while maintaining reasonable performance characteristics. Based on these results, we select $n = 8$ as the optimal parameter for subsequent experiments. This choice balances proof generation efficiency with the practical requirement of processing multiple messages per communication session, while minimizing the impact on MCP communication performance.

\subsection{Experiment 2: MCP Communication Performance Impact}

The second experiment evaluates the impact of zk-MCP on standard MCP communication performance. This experiment is critical for demonstrating that zk-MCP maintains the efficiency of standard MCP while adding privacy-preserving audit capabilities. According to the zk-MCP protocol specification in Sec.~4.4, zero-knowledge proof generation is performed \textit{asynchronously} after MCP communication sessions complete, 
ensuring that MCP message exchange latency remains unchanged.
\begin{figure*}[!ht]
    \centering
    \subfloat[DeepSeek V3]{\includegraphics[trim=0 0 0 300, clip, width=0.9648\textwidth]{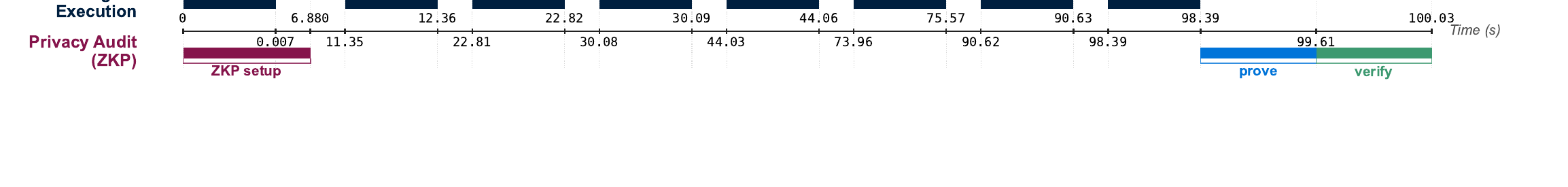}\label{fig:mcp_performance_ds}}\\
    \vspace{0.2cm}
    \subfloat[GPT-4.1mini]{\includegraphics[trim=0 0 0 300, clip, width=0.9648\textwidth]{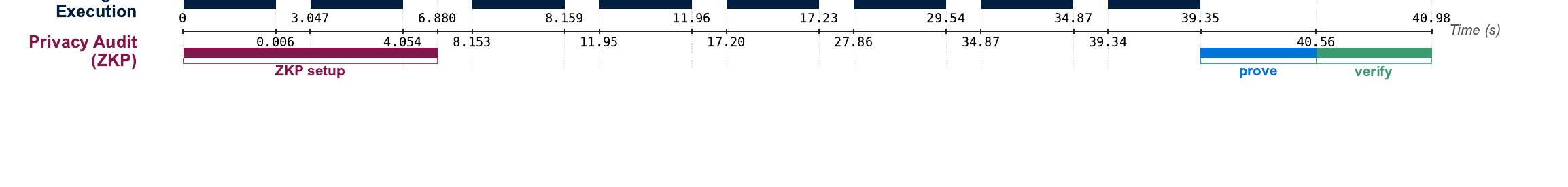}\label{fig:mcp_performance_4o}}\\
    \vspace{0.2cm}
    \subfloat[GPT-3.5 turbo]{\includegraphics[trim=0 0 0 300, clip, width=0.9648\textwidth]{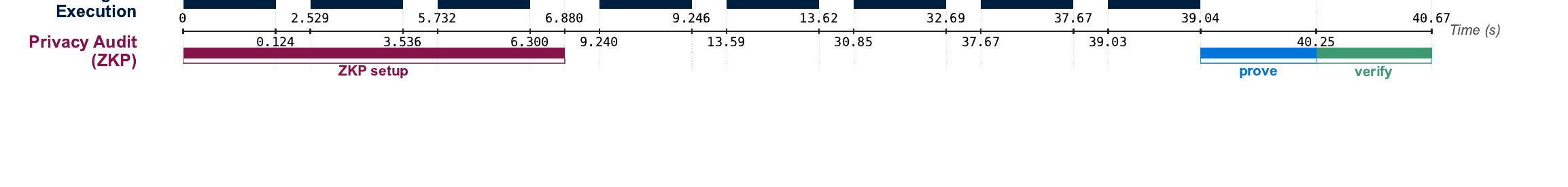}\label{fig:mcp_performance_35}}
    \caption{Performance comparison of standard MCP communication versus zk-MCP-enhanced communication across three LLMs: (a) DeepSeek V3, (b) GPT-4.1mini, and (c) GPT-3.5 turbo. Each subfigure shows the timing breakdown including MCP communication time, proof generation time, and total session time, demonstrating that zk-MCP introduces less than 4.14\% overhead on original MCP communication performance.}
    \label{fig:mcp_performance}
    \end{figure*}

The experiment is conducted using three different LLMs as agents: DeepSeek V3, GPT-4.1mini, and GPT-3.5 turbo. Each AI agent executes a series of MCP tool calls through specific prompts, while network packet capture is used to record communication timing. We measure the following metrics: (1) MCP message exchange latency (baseline standard MCP performance); (2) total session time including proof generation; (3) proof generation overhead as a percentage of total communication time; and (4) the time distribution across Setup, Prove, and Verify phases.


The experimental protocol follows the multi-session communication protocol described in Algorithm~\ref{alg:zk_mcp_protocol}. During the MCP operation phase, agents exchange messages through standard MCP protocol without any modification. Message parsing and storage occur asynchronously in parallel with MCP communication, ensuring non-blocking operation. After the MCP session completes, agents generate zero-knowledge proofs using the collected message metadata.

Figure~\ref{fig:mcp_performance} presents the experimental results comparing standard MCP communication with zk-MCP-enhanced communication across the three LLMs. As shown in Fig.~\ref{fig:mcp_performance}(a) for DeepSeek V3, Fig.~\ref{fig:mcp_performance}(b) for GPT-4.1mini, and Fig.~\ref{fig:mcp_performance}(c) for GPT-3.5 turbo, the results consistently demonstrate that zk-MCP introduces less than 4.14\% overhead on original MCP communication performance across all three models. This minimal overhead is achieved through the asynchronous proof generation design, where:

\begin{itemize}
    \item \textbf{Setup Phase}: The Setup process is independent of message content and can be performed once at protocol initialization, as specified in Sec.~4.4. The Setup time does not affect MCP communication latency since it occurs before communication sessions begin.
    
    \item \textbf{Communication Phase}: MCP message exchange proceeds with standard MCP performance, as zero-knowledge proof generation is completely decoupled from the communication process. Message parsing and storage occur in parallel with MCP communication, with $O(1)$ time complexity per message.
    
    \item \textbf{Proof Generation}: The Prove phase is executed \textit{after} all MCP tool calls complete, ensuring that MCP communication timing is not affected. The proof generation overhead is independent of communication timing and does not block subsequent communication sessions.
    
    \item \textbf{Verification Phase}: The Verify phase is performed by the ASP immediately after the zero-knowledge proof is submitted. The verification time accounts for only a small proportion of the total time, specifically 4.14\%.
\end{itemize}

The experimental data across all three LLMs shows consistent results. For a typical communication session with 8 messages, the MCP communication time remains at baseline performance for each model, while proof generation adds minimal overhead. Specifically, as illustrated in Fig.~\ref{fig:mcp_performance}, DeepSeek V3, GPT-4.1mini, and GPT-3.5 turbo all demonstrate that zk-MCP introduces less than 4.14\% overhead on their respective baseline MCP communication performance. This consistent result across different LLMs validates the design principle stated in Sec.~4.5: zk-MCP enables verifiable audit capabilities while maintaining full compatibility with standard MCP communication efficiency regardless of the underlying AI agent implementation.

The experimental results demonstrate that while proof generation introduces overhead, the asynchronous design ensures that this overhead has minimal impact on MCP communication performance, with total overhead remaining below 4.14\% of communication costs. This suggests that zk-MCP is able to provide privacy-preserving audit verification with limited impact on standard MCP performance, indicating potential applicability to IoA frameworks where communication efficiency is important and various types of AI agents may need to be supported.

\section{Conclusion}
\label{sec:conclusion}

This paper presents the first zero-knowledge proof-based privacy-preserving audit framework for agent communications in IoA systems. By integrating MCP's context-aware communication capabilities with zk-SNARK's privacy-preserving verification mechanisms, zk-MCP enables verifiable audit trails without exposing sensitive communication content or compromising agent privacy, enabling mutual audit capabilities between agents while maintaining complete communication confidentiality.

Our theoretical analysis demonstrates that zk-MCP satisfies data authenticity and communication privacy properties. Through comprehensive experimental evaluation, we first establish the feasibility of using zero-knowledge proofs for audit verification by demonstrating circuit parameter scalability across different message counts. The experimental results show that while proof generation introduces overhead, the asynchronous design ensures that this overhead has minimal impact on MCP communication performance, with total overhead remaining below 4.14\% of communication costs. Our implementation, which utilizes Circom circuits and integrates with MCP's bidirectional communication, offers an initial exploration of the framework's practical feasibility for IoA deployments. While preliminary results are promising, further testing and refinement will be needed to fully assess its applicability to privacy-preserving audit verification in diverse and regulated environments.


\bibliographystyle{IEEEtran}
\bibliography{references}

\end{document}